\documentclass{elsart}
\usepackage{epsfig}

\begin{document}

\begin{frontmatter}

\title{Exponential distribution of financial returns
    at mesoscopic time lags: a new stylized fact}
  
  \author{A. Christian Silva, Richard E. Prange, Victor M.
    Yakovenko\thanksref{email}}

\thanks[email]{http://www2.physics.umd.edu/\~{}yakovenk/econophysics.html}
 
\address{Department of Physics, University of Maryland, College Park,
  MD 20742-4111, USA \hfill \bf cond-mat/0401225, v.1: 13 January 2004,
  v.2: 24 July 2004}

% Version J to cond-mat, edited by VMY on July 24, 2004

\begin{abstract}
  We study the probability distribution of stock returns at mesoscopic
  time lags (return horizons) ranging from about an hour to about a
  month.  While at shorter microscopic time lags the distribution has
  power-law tails, for mesoscopic times the bulk of the distribution
  (more than 99\% of the probability) follows an exponential law.  The
  slope of the exponential function is determined by the variance of
  returns, which increases proportionally to the time lag.  At longer
  times, the exponential law continuously evolves into Gaussian
  distribution.  The exponential-to-Gaussian crossover is well
  described by the analytical solution of the Heston model with
  stochastic volatility.
\end{abstract}

\begin{keyword}
  Econophysics \sep Exponential distribution \sep Stylized facts \sep
  Stochastic volatility \sep Heston model \sep Stock market returns
  \sep Empirical characteristic function
  \PACS 02.50.-r \sep 89.65.-s
\end{keyword}

\end{frontmatter}

%%%%%%%%%%%%%%%%%%%%%%%%%%%%%%%%%%%%%%%%%%%%%%%%%%%%%%%%%%%%%%%%%%%%%%
\section{Introduction}
%%%%%%%%%%%%%%%%%%%%%%%%%%%%%%%%%%%%%%%%%%%%%%%%%%%%%%%%%%%%%%%%%%%%%%

The empirical probability distribution functions (EDFs) for different
assets have been extensively studied by the econophysics community in
recent years
\cite{Bouchaud,Stanley1999a,Stanley1999b,DY,SY,India,Japan,Germany,Brazil,Miranda}.
Stock and stock-index returns have received special attention.  We
focus here on the EDFs of the returns of individual large American
companies from 1993 to 1999, a period without major market
disturbances.  By `return' we always mean `log-return', the difference
of the logarithms of prices at two times separated by a time lag $t$.

The time lag $t$ is an important parameter: the EDFs evolve with this
parameter. At micro lags (typically shorter than one hour), effects
such as the discreteness of prices and transaction times, correlations
between successive transactions, and fluctuations in trading rates
become important.  Power-law tails of EDFs in this regime have been
much discussed in the literature before
\cite{Stanley1999a,Stanley1999b}.  At `meso' time lags (typically from
an hour to a month), continuum approximations can be made, and some
sort of diffusion process is plausible, eventually leading to a normal
Gaussian distribution.  On the other hand, at `macro' time lags, the
changes in the mean market drifts and macroeconomic `convection'
effects can become important, so simple results are less likely to be
obtained.  The boundaries between these domains to an extent depend on
the stock, the market where it is traded, and the epoch. The
micro-meso boundary can be defined as the time lag above which
power-law tails constitute a very small part of the EDF. The
meso-macro boundary is more tentative, since statistical data at long
time lags become sparse.

The first result is that we extend to meso time lags a stylized fact
known since the 19th century \cite{Regnault} (quoted in \cite{Taqqu}):
with a careful definition of time lag $t$, the variance of returns is
proportional to $t$.

The second result is that log-linear plots of the EDFs show prominent
straight-line (tent-shape) character, i.e.\ the bulk (about 99\%) of
the probability distribution of log-return follows an exponential law.
The exponential law applies to the central part of EDFs, i.e.\ not too
big log-returns.  For the far tails of EDFs, usually associated with
power laws at micro time lags, we do not have enough statistically
reliable data points at meso lags to make a definite conclusion.
Exponential distributions have been reported for some world markets
\cite{DY,SY,India,Japan,Germany,Brazil,Miranda} and briefly mentioned
in the book \cite{Bouchaud} (see Fig.\ 2.12).  However, the
exponential law has not yet achieved the status of a stylized fact.
Perhaps this is because influential work
\cite{Stanley1999a,Stanley1999b} has been interpreted as finding that
the individual returns of all the major US stocks for micro to macro
time lags have the same power law EDFs, if they are rescaled by the
volatility.

The Heston model is a plausible diffusion model with stochastic
volatility, which reproduces the timelag-variance proportionality and
the cross\-over from exponential distribution to Gaussian.  This model
was first introduced by Heston, who studied option prices
\cite{Heston}. Later Dr\u{a}gulescu and Yako\-ven\-ko (DY) derived a
convenient closed-form expression for the probability distribution of
returns in this model and applied it to stock indexes from 1 day to 1
year \cite{DY}.  The third result is that the DY formula with three
lag-independent parameters reasonably fits the time evolution of EDFs
at meso lags.

%%%%%%%%%%%%%%%%%%%%%%%%%%%%%%%%%%%%%%%%%%%%%%%%%%%%%%%%%%%%%%%%%%%%%%
\section{Probability distribution of log-returns in the Heston model}
%%%%%%%%%%%%%%%%%%%%%%%%%%%%%%%%%%%%%%%%%%%%%%%%%%%%%%%%%%%%%%%%%%%%%%

In this section, the Heston model \cite{Heston} and the DY formula
\cite{DY} are briefly summarized.  The price $S_t$ of a model stock
obeys the stochastic differential equation of multiplicative Brownian
motion: $dS_t = \mu S_t\, dt + \sqrt{v_t} S_t\, dW_t^{(1)}$.  Here the
subscript $t$ indicates time dependence, $\mu$ is the drift parameter,
$W_t^{(1)}$ is a standard random Wiener process, and $v_t$ is the
fluctuating time-dependent variance.  The detrended log-return is
defined as $x_t=\ln(S_t/S_0)-\mu t$, although detrending is a minor
correction at meso lags.  In the Heston model, the variance $v_t$
obeys the mean-reverting stochastic differential equation:
\begin{equation} \label{eqVar}
  dv_t = -\gamma(v_t - \theta)\,dt + \kappa\sqrt{v_t}\,dW_t^{(2)}.
\end{equation}
Here $\theta$ is the long-time mean of $v$, $\gamma$ is the rate of
relaxation to this mean, and $\kappa$ is the variance noise.  We take
the Wiener processes $W_t^{(1,2)}$ to be uncorrelated.

The DY formula \cite{DY} for the probability density function (PDF)
$P_t(x)$ is:
\begin{eqnarray}
  && P_t(x) =
  \int\limits_{-\infty}^{+\infty} \frac{dk}{2\pi}\,
  e^{ikx + F_{\tilde t}(k)},\;
  F_{\tilde t}(k)=\frac{\alpha\tilde t}{2}
  - \alpha\ln\left[\cosh\frac{\Omega\tilde t}{2} +
  \frac{\Omega^2+1}{2\Omega}
  \sinh\frac{\Omega\tilde t}{2}\right],
\label{eq:DY}  \\
  && \tilde t=\gamma t, \quad \alpha=2\gamma\theta/\kappa^2,
  \quad \Omega=\sqrt{1+(k\kappa/\gamma)^2}, \quad 
  \sigma_t^2\equiv\langle x_t^2\rangle=\theta t.
\label{eq:DY2}
\end{eqnarray}
The variance $\sigma_t^2\equiv\langle x_t^2\rangle$ (\ref{eq:DY2}) of
the PDF (\ref{eq:DY}) increases linearly in time, while $\langle
x_t\rangle=0$.  The three parameters of the model are $\gamma$,
$\theta$ and $\alpha$.  At short and long time lags, the PDF
(\ref{eq:DY}) reduces to exponential (if $\alpha = 1$) and Gaussian
\cite{DY}:
\begin{equation}
  P_{t}(x)\propto\left\{
    \begin{array}{ll}
    \exp(-|x|\sqrt{2/\theta t}),  & \quad \tilde t =\gamma t\ll1, \\
    \exp(-x^{2}/2 \theta t), &  \quad \tilde t=\gamma t\gg1.
    \end{array}
    \right.
\label{short-long}
\end{equation}
In both limits, it scales with the volatility:
$P_{t}(x)=f(x/\sqrt{\langle x_t^2\rangle})=f(x/\sqrt{\theta t})$,
where $f$ is the exponential or the Gaussian function.

%%%%%%%%%%%%%%%%%%%%%%%%%%%%%%%%%%%%%%%%%%%%%%%%%%%%%%%%%%%%%%%%%%%%%%%
\section{Data analysis and discussion}
%%%%%%%%%%%%%%%%%%%%%%%%%%%%%%%%%%%%%%%%%%%%%%%%%%%%%%%%%%%%%%%%%%%%%%%

We analyzed the data from Jan/1993 to Jan/2000 for $27$ Dow companies,
but show results only for four large cap companies: Intel (INTC) and
Microsoft (MSFT) traded at NASDAQ, and IBM and Merck (MRK) traded at
NYSE.  We use two databases, TAQ to construct the intraday returns and
Yahoo database for the interday returns. The intraday time lags were
chosen at multiples of 5 minutes, which divide exactly the 6.5 hours
(390 minutes) of the trading day. The interday returns are as
described in \cite{DY,SY} for time lags from 1 day to 1 month = 20
trading days.

%%%%%%%%%%%%%%%%%%%%%%%%%%%%%%%%%%%%%%%%%%%%%%%%%%%%%%%%%%%%%%%%%%%%%%
\begin{figure}
\centerline{
\epsfig{file=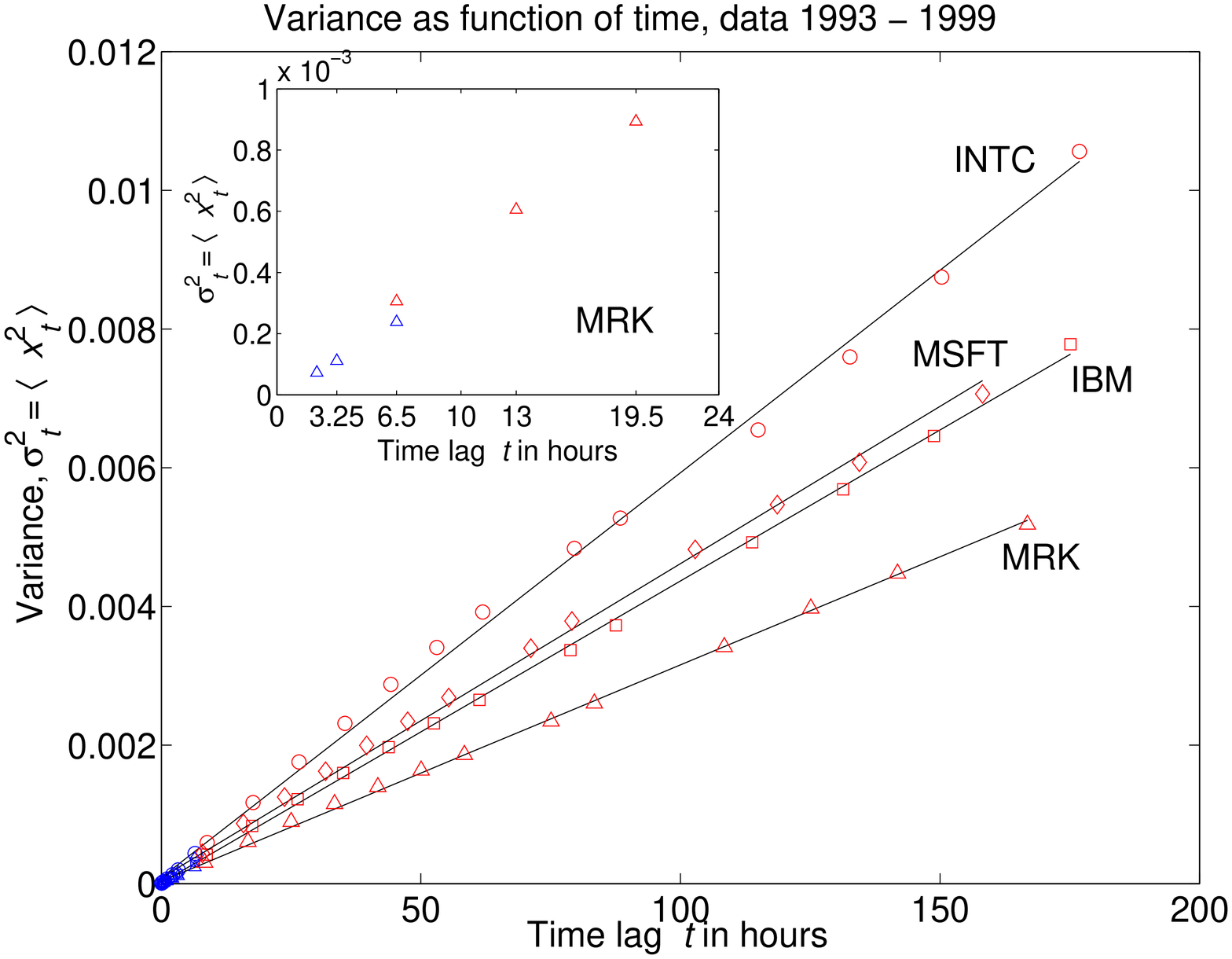,width=0.41\linewidth,angle=-90}
\epsfig{file=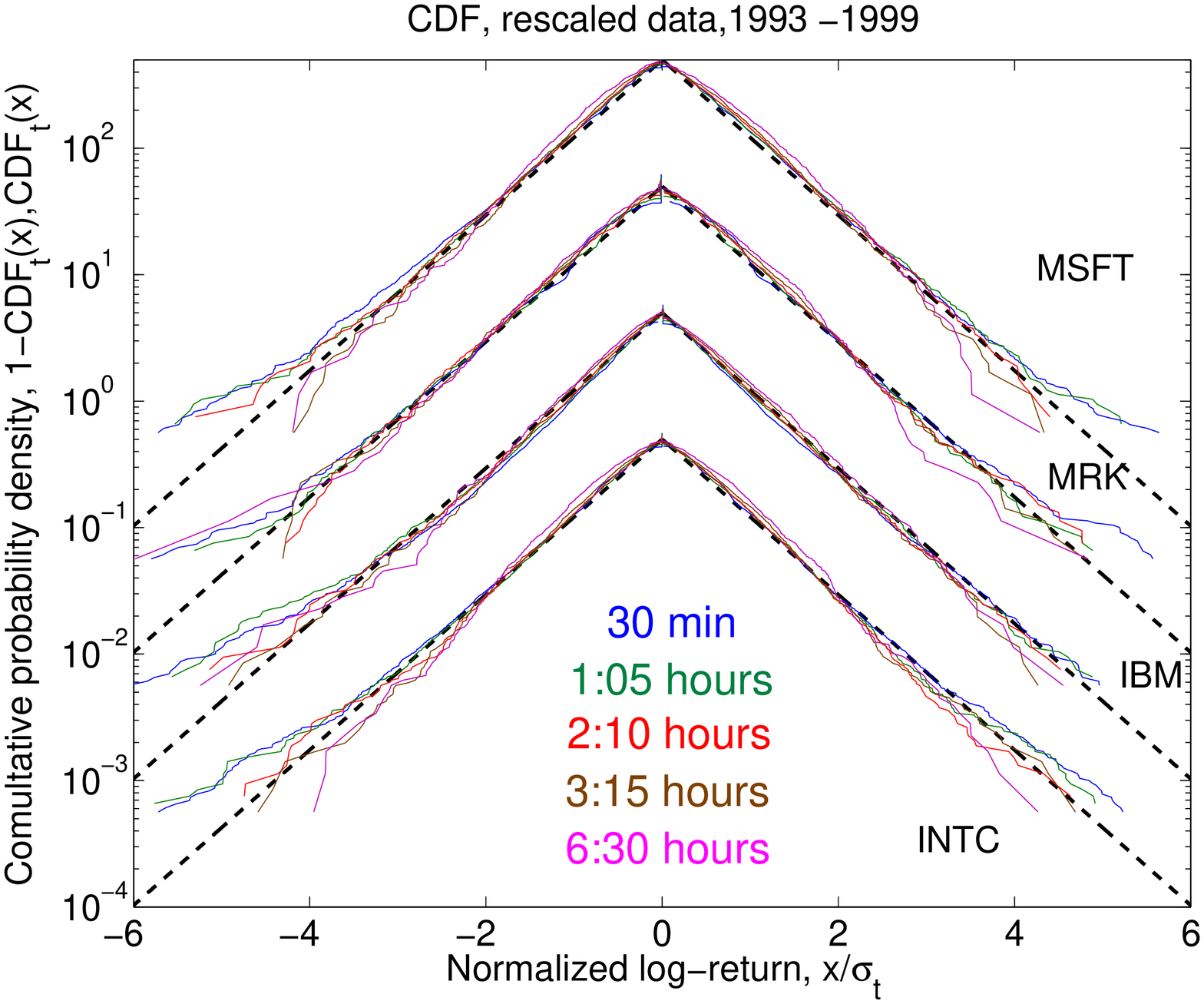,width=0.41\linewidth,angle=-90}}
\caption{\footnotesize\sf Left panel: Variance
  $\langle x_t^2\rangle$ vs.\ time lag $t$. Solid lines: Linear fits
  $\langle x_t^2\rangle=\theta t$. Inset: Variances for MRK before
  adjustment for the effective overnight time $T_n$. Right panel:
  Log-linear plots of CDFs vs.\ $x/\sqrt{\theta t}$.  Straight dashed
  lines $-|x|\sqrt{2/\theta t}$ are predicted by the DY formula
  (\ref{short-long}) in the short-time limit.  The curves are offset
  by a factor of 10.}
\label{fig:Var}
\end{figure}
%%%%%%%%%%%%%%%%%%%%%%%%%%%%%%%%%%%%%%%%%%%%%%%%%%%%%%%%%%%%%%%%%%

%%%%%%%%%%%%%%%%%%%%%%%%%%%%%%%%%%%%%%%%%%%%%%%%%%%%%%%%%%%%%%%%%%%%%%
\begin{figure}[b]
\centerline{
  \epsfig{file=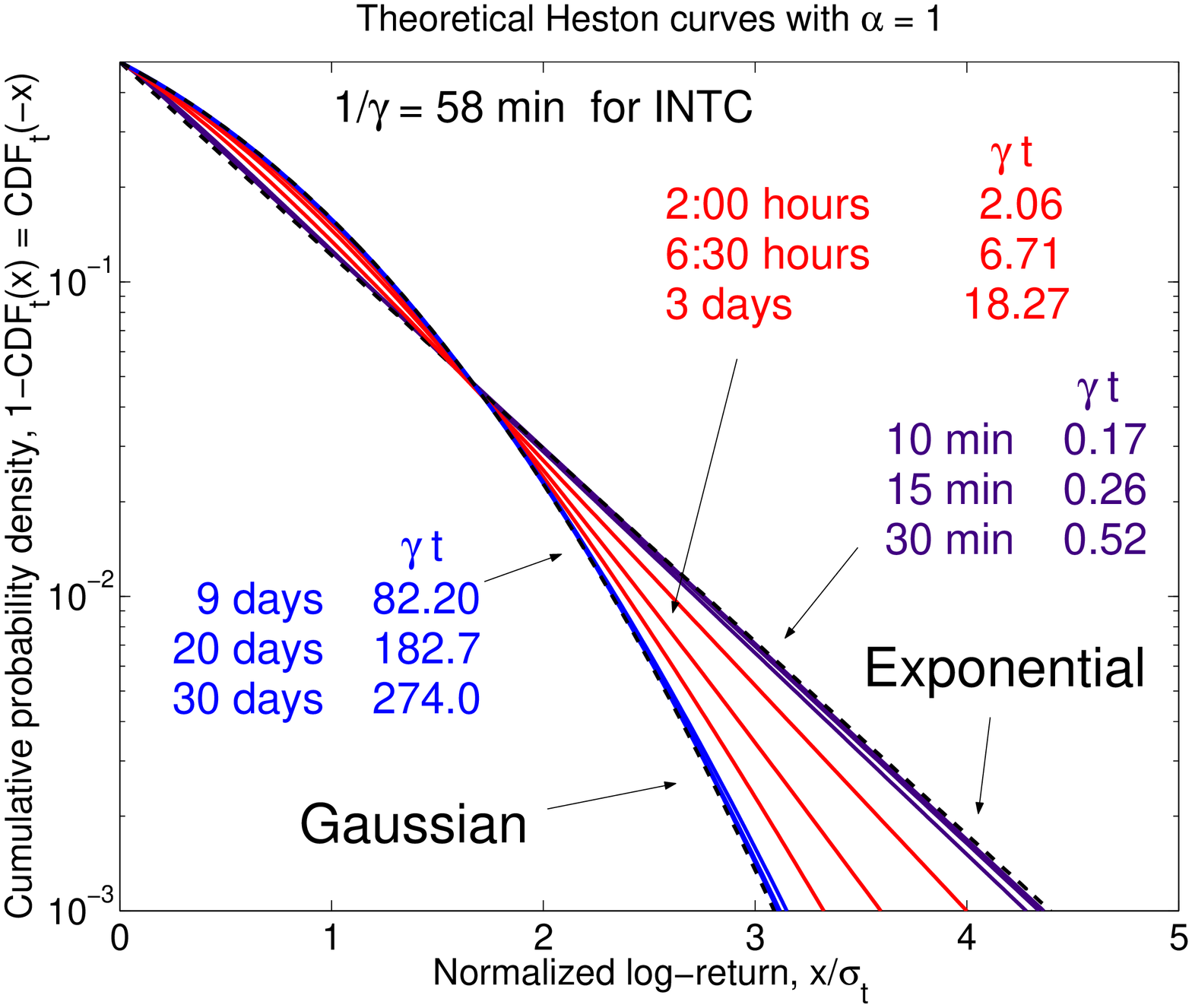,width=0.41\linewidth,angle=-90}
  \epsfig{file=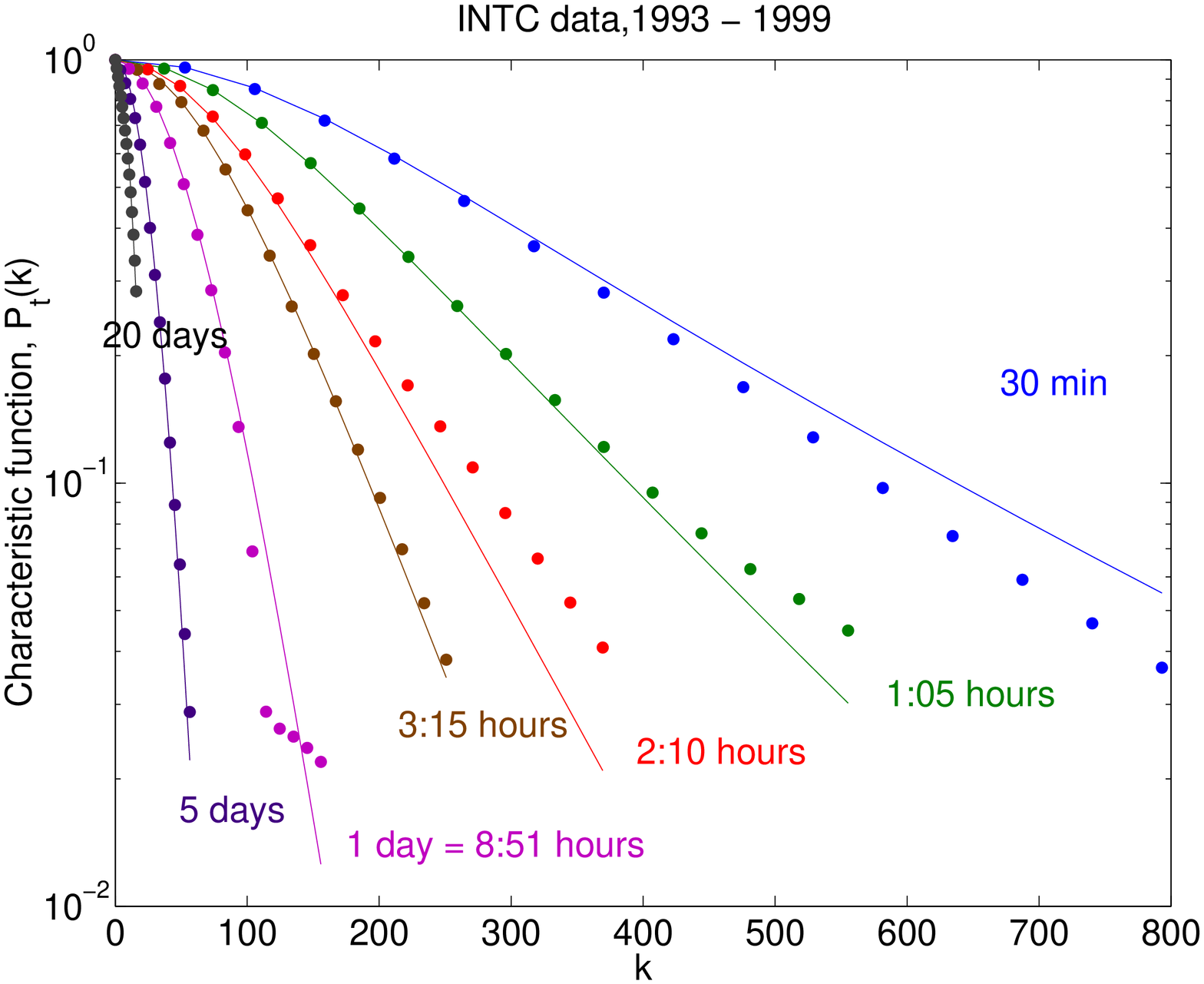,width=0.41\linewidth,angle=-90}}
\caption{\footnotesize\sf Left panel: Theoretical CDFs for the Heston
  model plotted vs.\ $x/\sqrt{\theta t}$.  The curves interpolate
  between the short-time exponential and long-time Gaussian scalings.
  Right panel: Comparison between empirical (points) and the DY
  theoretical (curves) characteristic functions $\tilde P_t(k)$.}
\label{fig:ThPk}
\end{figure}
%%%%%%%%%%%%%%%%%%%%%%%%%%%%%%%%%%%%%%%%%%%%%%%%%%%%%%%%%%%%%%%%%%%%%%

In order to connect the interday and intraday data, we have to
introduce an effective overnight time lag $T_n$.  Without this
correction, the open-to-close and close-to-close variances would have
a discontinuous jump at 1 day, as shown in the inset of the left panel
of Fig.\ \ref{fig:Var}.  By taking the open-to-close time to be 6.5
hours, and the close-to-close time to be 6.5 hours + $T_n$, we find
that variance $\langle x_t^2\rangle$ is proportional to time $t$, as
shown in the left panel of Fig.\ \ref{fig:Var}.  The slope gives us
the Heston parameter $\theta$ in Eq.\ (\ref{eq:DY2}).  $T_n$ is about
2 hours (see Table \ref{Parameters}).

In the right panel of Fig.\ \ref{fig:Var}, we show the log-linear
plots of the cumulative distribution functions (CDFs) vs.\ normalized
return $x/\sqrt{\theta t}$.  The $\makebox{CDF}_t(x)$ is defined as
$\int_{-\infty}^xP_t(x')\,dx'$, and we show $\makebox{CDF}_t(x)$ for
$x<0$ and $1-\makebox{CDF}_t(x)$ for $x>0$.  We observe that CDFs for
different time lags $t$ collapse on a single straight line without any
further fitting (the parameter $\theta$ is taken from the fit in the
left panel).  More than 99\% of the probability in the central part of
the tent-shape distribution function is well described by the
exponential function.  Moreover, the collapsed CDF curves agree with
the DY formula (\ref{short-long})
$P_{t}(x)\propto\exp(-|x|\sqrt{2/\theta t})$ in the short-time limit
for $\alpha=1$ \cite{DY}, which is shown by the dashed lines.

%%%%%%%%%%%%%%%%%%%%%%%%%%%%%%%%%%%%%%%%%%%%%%%%%%%%%%%%%%%%%%%%%%%
\begin{table}[b]
\caption{\footnotesize\sf Fitting parameters of the Heston model
  with $\alpha=1$ for the 1993--1999 data.
  \label{Parameters}}
\begin{tabular}{c|cccccccc}
\hline
& $\gamma$ & $1/\gamma$ & $\theta$ & $\mu$ & $T_{n}$ \\
& ${1\over{\rm hour}}$ & hour & ${1\over{\rm year}}$ 
& ${1\over{\rm year}}$ & hour \\
\hline
INTC & $1.029$ & $0 \colon 58$ & $13.04\%$ & $39.8\%$ & $2 \colon 21$ \\
\hline
IBM & $0.096$ & $10 \colon 25$ & $9.63\%$ & $35.3\%$ & $2 \colon 16$  \\
\hline
MRK & $0.554$ & $1 \colon 48$ & $6.57\%$ & $29.4\%$ & $1 \colon 51$ \\
\hline
MSFT & $1.284$ & $0 \colon 47$ & $9.06\%$ & $48.3\%$ & $1 \colon 25$ \\
\hline
\end{tabular}
\end{table}
%%%%%%%%%%%%%%%%%%%%%%%%%%%%%%%%%%%%%%%%%%%%%%%%%%%%%%%%%%%%%%%%%%%

Because the parameter $\gamma$ drops out of the asymptotic Eq.\ 
(\ref{short-long}), it can be determined only from the crossover
regime between short and long times, which is illustrated in the left
panel of Fig.\ \ref{fig:ThPk}.  We determine $\gamma$ by fitting the
characteristic function $\tilde P_t(k)$, a Fourier transform of
$P_t(x)$ with respect to $x$.  The theoretical characteristic function
of the Heston model is $\tilde P_t(k)=e^{F_{\tilde t}(k)}$
(\ref{eq:DY}).  The empirical characteristic functions (ECFs) can be
constructed from the data series by taking the sum $\tilde P_t(k)={\rm
  Re}\sum_{x_t}\exp(-ikx_t)$ over all returns $x_t$ for a given $t$
\cite{BookCh}.  Fits of ECFs to the DY formula (\ref{eq:DY}) are shown
in the right panel of Fig.\ \ref{fig:ThPk}.  The parameters determined
from the fits are given in Table \ref{Parameters}.

%%%%%%%%%%%%%%%%%%%%%%%%%%%%%%%%%%%%%%%%%%%%%%%%%%%%%%%%%%%%%%%%%%%%%%
\begin{figure}
\centerline{
  \epsfig{file=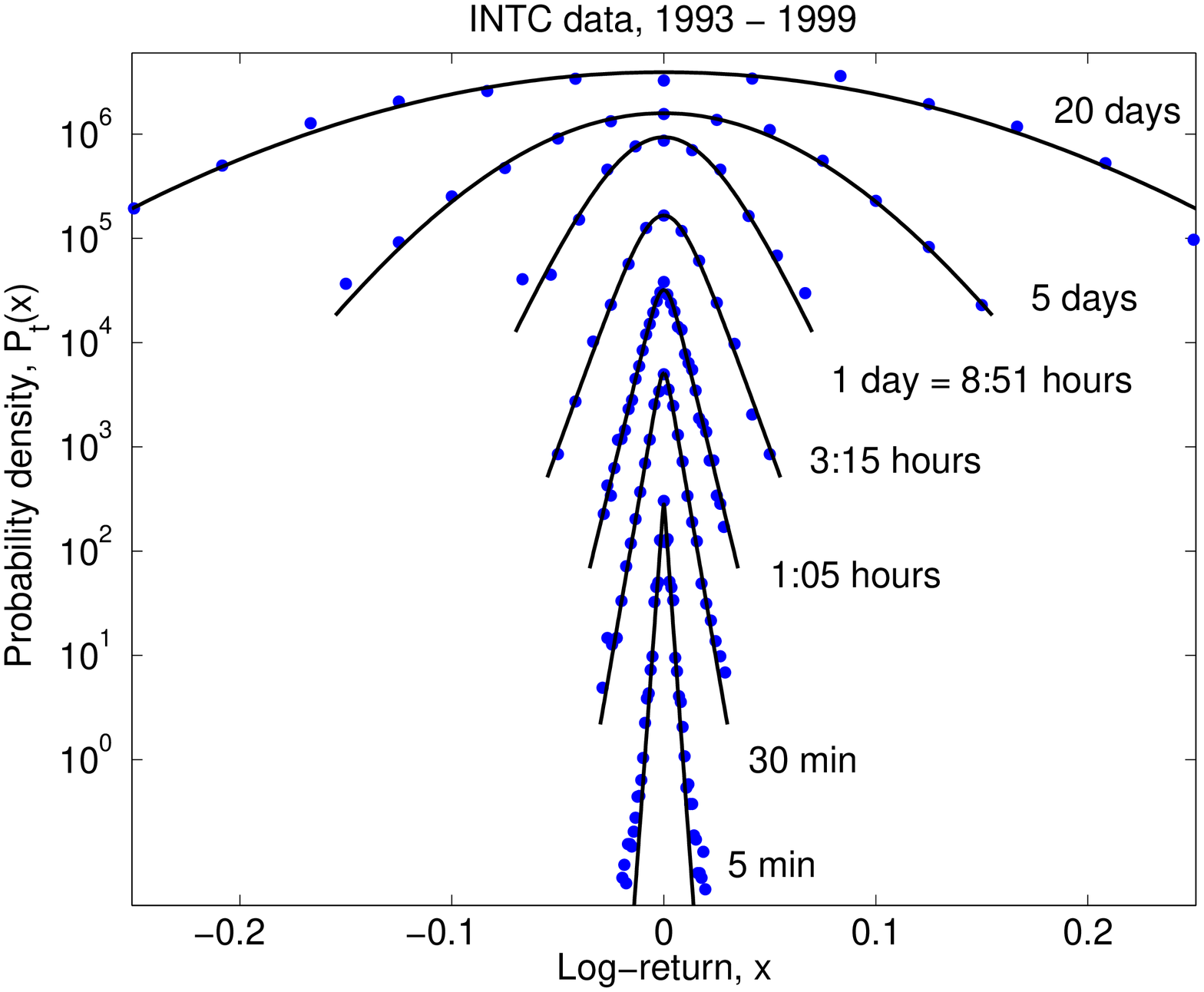,width=0.41\linewidth,angle=-90}
  \epsfig{file=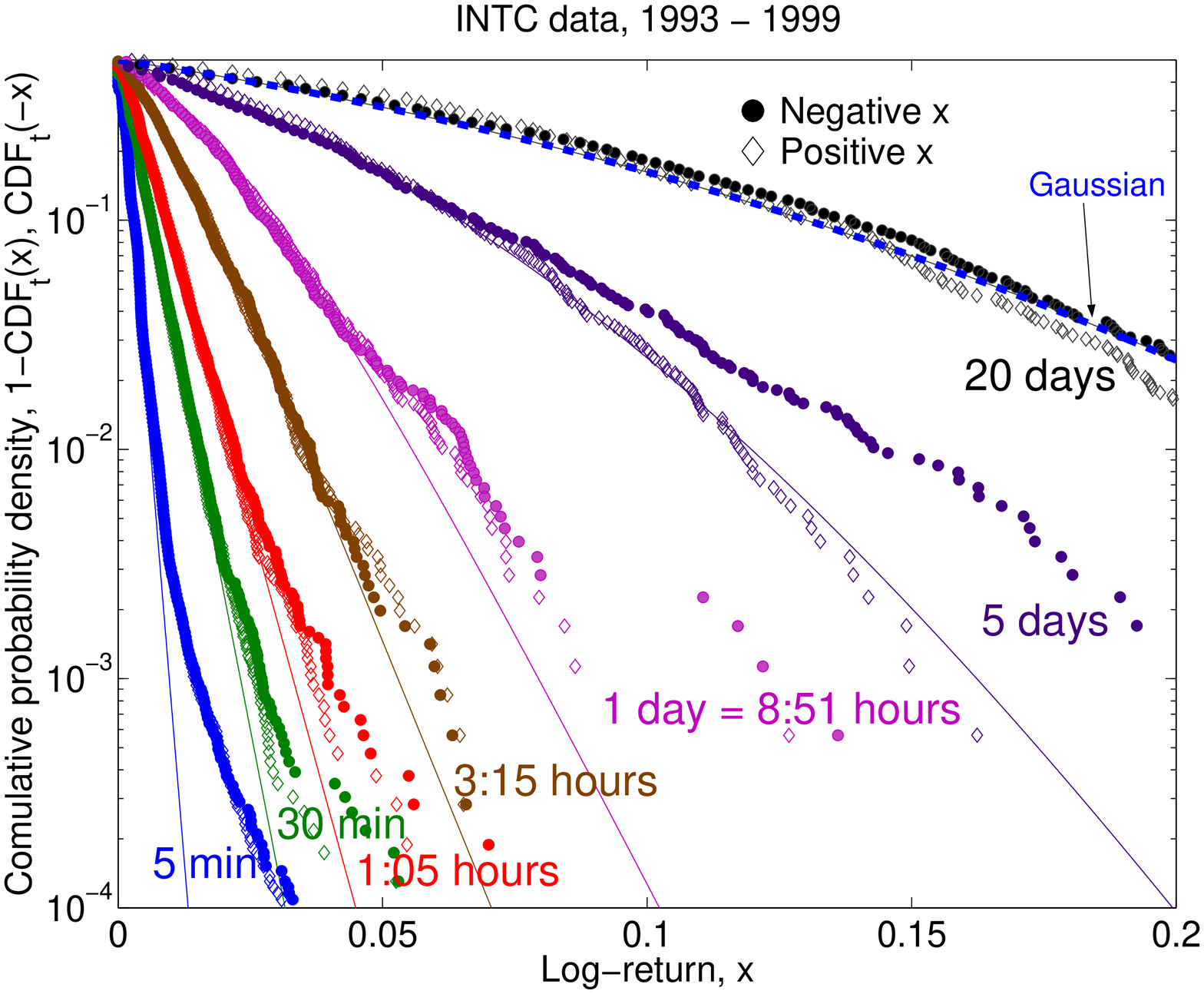,width=0.41\linewidth,angle=-90}}
\caption{\footnotesize\sf Comparison between the 1993--1999 Intel
  data (points) and the DY formula (\ref{eq:DY}) (curves) for PDF
  (left panel) and CDF (right panel).}
\label{fig:PdfCdf}
\end{figure}
%%%%%%%%%%%%%%%%%%%%%%%%%%%%%%%%%%%%%%%%%%%%%%%%%%%%%%%%%%%%%%%%%%%%%%%

\begin{figure}[b]
\centerline{
  \epsfig{file=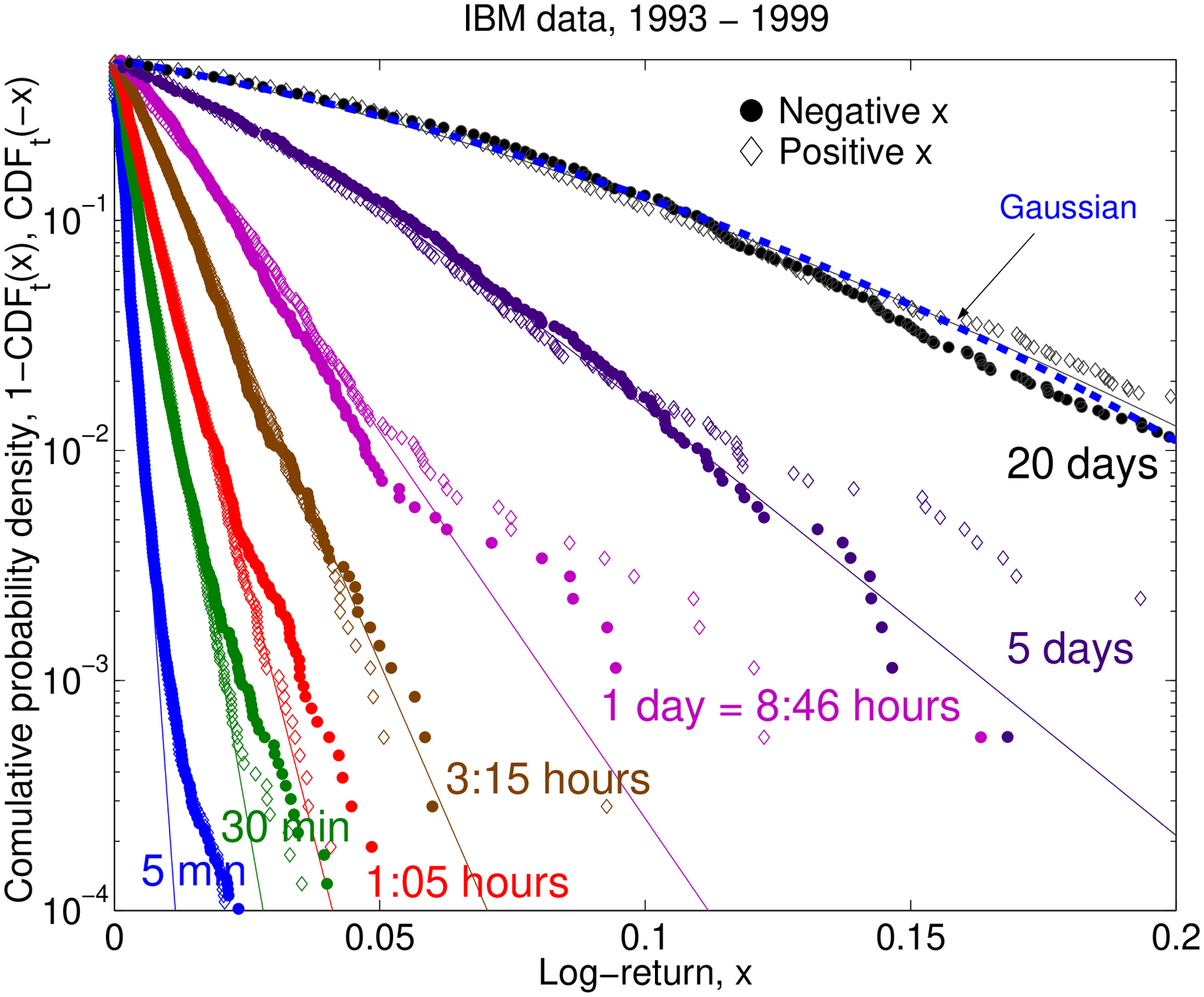,width=0.41\linewidth,angle=-90}
  \epsfig{file=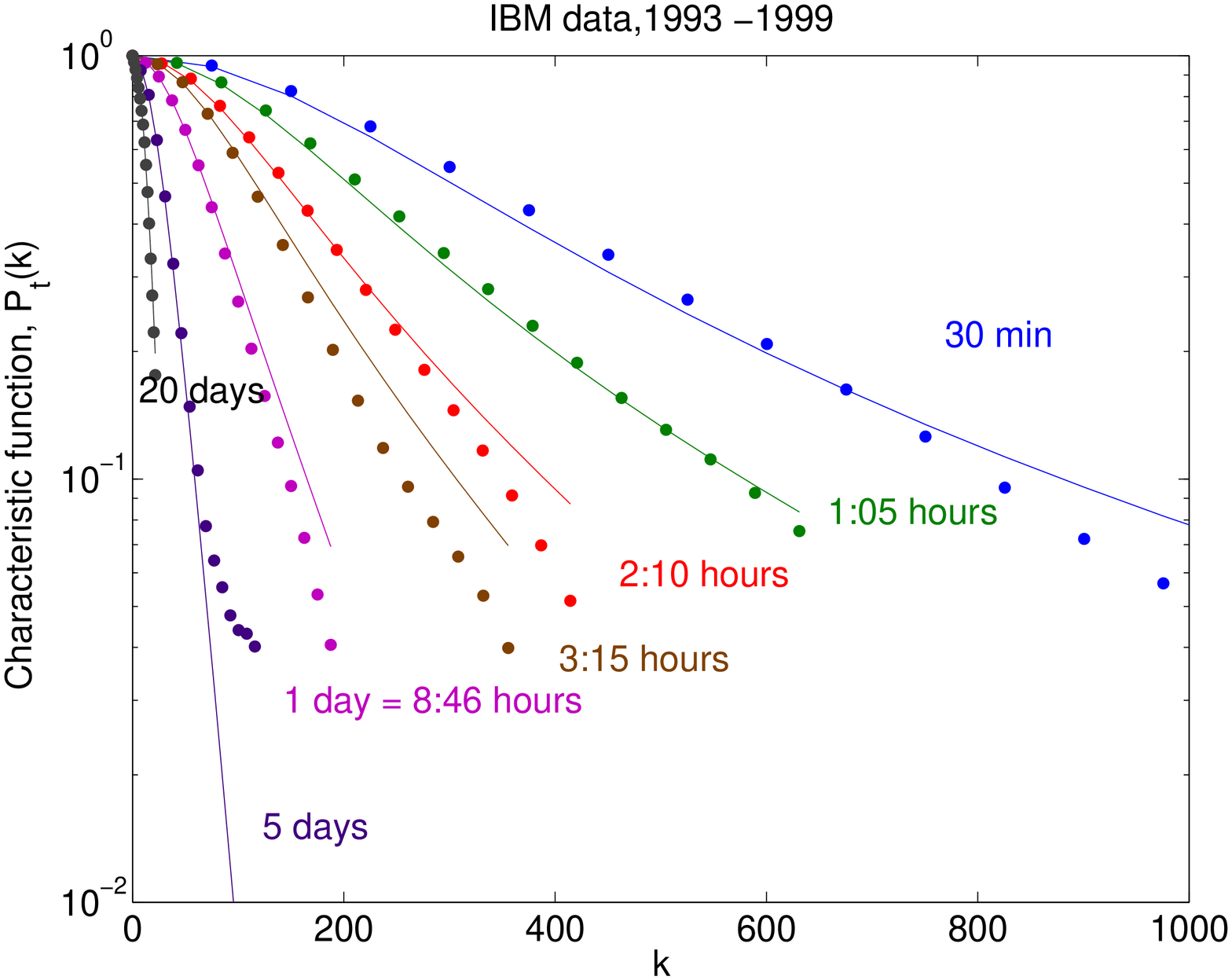,width=0.41\linewidth,angle=-90}}
\centerline{
  \epsfig{file=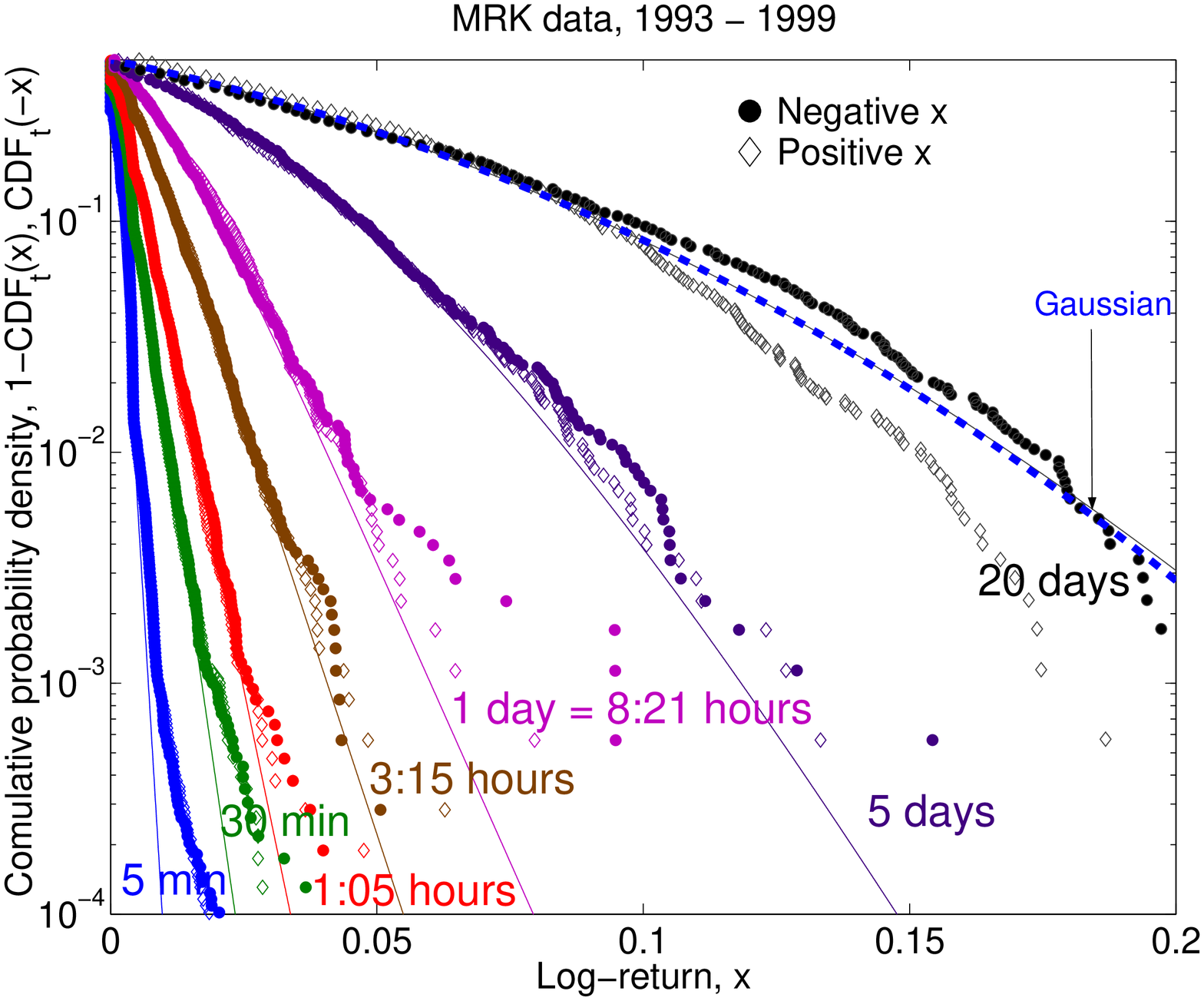,width=0.41\linewidth,angle=-90}
  \epsfig{file=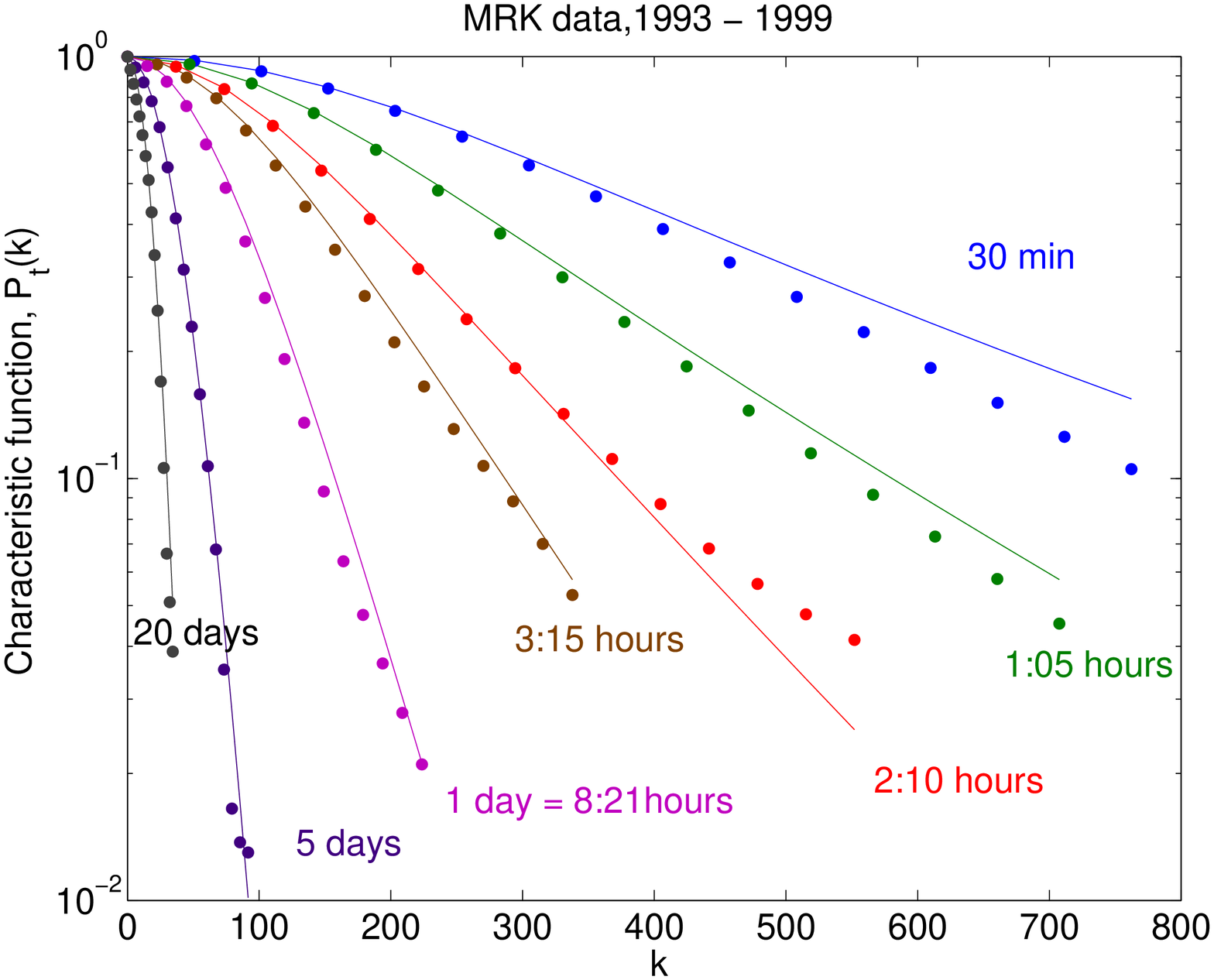,width=0.41\linewidth,angle=-90}}
\centerline{
  \epsfig{file=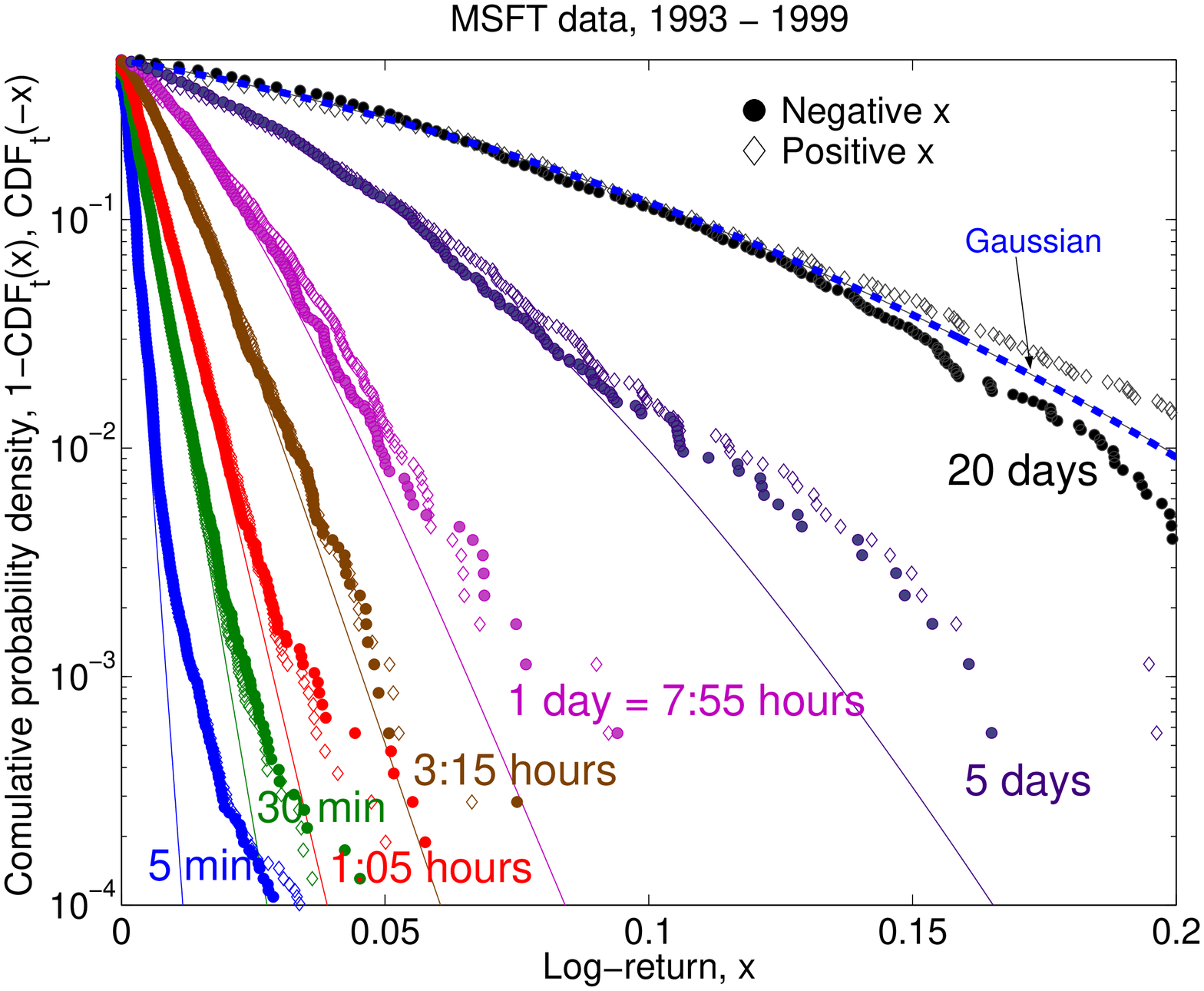,width=0.41\linewidth,angle=-90}
  \epsfig{file=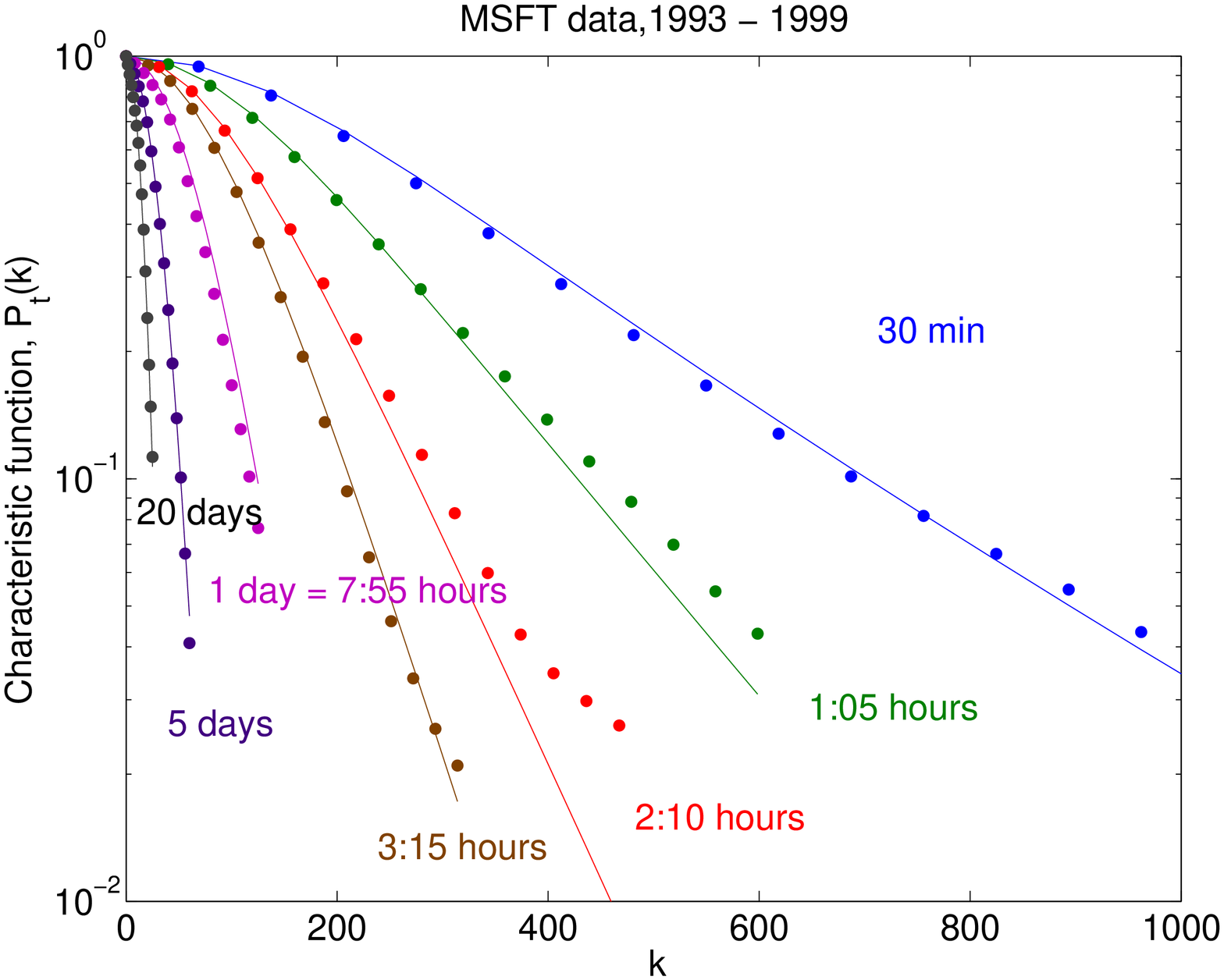,width=0.41\linewidth,angle=-90}}
\caption{\footnotesize\sf Comparison between empirical data
  (symbols) and the DY formula (\ref{eq:DY}) (lines) for CDF (left
  panels) and characteristic function (right panels).}
  \label{fig:many}
\end{figure}

In the left panel of Fig.\ \ref{fig:PdfCdf} we compare the empirical
PDF $P_t(x)$ with the DY formula (\ref{eq:DY}).  The agreement is
quite good, except for the very short time lag of 5 minutes, where the
tails are visibly fatter than exponential.  In order to make a more
detailed comparison, we show the empirical CDFs (points) with the
theoretical DY formula (lines) in the right panel of Fig.\ 
\ref{fig:PdfCdf}.  We see that, for micro time lags of the order of 5
minutes, the power-law tails are significant.  However, for meso time
lags, the CDFs fall onto straight lines in the log-linear plot,
indicating exponential law.  For even longer time lags, they evolve
into the Gaussian distribution in agreement with the DY formula
(\ref{eq:DY}) for the Heston model.  To illustrate the point further,
we compare empirical and theoretical data for several other companies
in Fig.\ \ref{fig:many}.

In the empirical CDF plots, we actually show the ranking plots of
log-returns $x_t$ for a given $t$.  So, each point in the plot
represents a single instance of price change.  Thus, the last one or
two dozens of the points at the far tail of each plot constitute a
statistically small group and show large amount of noise.
Statistically reliable conclusions can be made only about the central
part of the distribution, where the points are dense, but not about
the far tails.

%%%%%%%%%%%%%%%%%%%%%%%%%%%%%%%%%%%%%%%%%%%%%%%%%%%%%%%%%%%%%%%%%%
\section{Conclusions}
\label{sec:conclusions}
%%%%%%%%%%%%%%%%%%%%%%%%%%%%%%%%%%%%%%%%%%%%%%%%%%%%%%%%%%%%%%%%%%

We have shown that in the mesoscopic range of time lags, the
probability distribution of financial returns interpolates between
exponential and Gaussian law.  The time range where the distribution
is exponential depends on a particular company, but it is typically
between an hour and few days. Similar exponential distributions have
been reported for the Indian \cite{India}, Japanese \cite{Japan},
German \cite{Germany}, and Brazilian markets \cite{Brazil,Miranda}, as
well as for the US market \cite{DY,SY} (see also Fig. 2.12 in
\cite{Bouchaud}).  The DY formula \cite{DY} for the Heston model
\cite{Heston} captures the main features of the probability
distribution of returns from an hour to a month with a single set of
parameters.  We believe that econophysicists should be aware of the
presence of the exponential distribution and recognize it as another
``stylized fact'' in the set of analytical tools for financial data
analysis.

We thank Chuck Lahaie from the Robert H. Smith School of Business at
UMD for help with the TAQ database.

%%%%%%%%%%%%%%%%%%%%%%%%%%%%%%%%%%%%%%%%%%%%%%%%%%%%%%%%%%%%%%%%%%%%%%

%%%%%%%%%%%%%%%%%%%%%%%%%%%%%%%%%%%%%%%%%%%%%%%%%%%%%%%%%%%%%%%%%%%%%%
%%%%%%%%%%%%%%%%%%%%%%%%%%%%%%%%%%%%%%%%%%%%%%%%%%%%%%%%%%%%%%%%%%%%%%%

\end{document}